\documentclass[journal=jacsat,manuscript=article]{achemso}
\setkeys{acs}{doi = true}
\usepackage{moreverb}
\usepackage{graphicx}
\usepackage{dcolumn}
\usepackage{bm}
\usepackage[version=3]{mhchem} 
\usepackage{color, soul}
\usepackage{comment}
\usepackage{booktabs}
\usepackage{makecell}

\author{Changbin Im}
\affiliation[Ulm University]
{Institute of Electrochemistry, Ulm University, Albert-Einstein-Allee 47, 89081 Ulm, Germany}

\author{Björn Kirchhoff}
\affiliation[Ulm University]
{Institute of Electrochemistry, Ulm University, Albert-Einstein-Allee 47, 89081 Ulm, Germany}

\author{Dariusz Mitoraj}
\affiliation[Ulm University]
{Institute of Electrochemistry, Ulm University, Albert-Einstein-Allee 47, 89081 Ulm, Germany}

\author{Igor Krivtsov}
\affiliation[University of Oviedo]
{Department of Chemical and Environmental Engineering, University of Oviedo, 33006 Oviedo, Spain}

\author{Attila Farkas}
\affiliation[Ulm University]
{Institute of Electrochemistry, Ulm University, Albert-Einstein-Allee 47, 89081 Ulm, Germany}

\author{Radim Beranek}
\affiliation[Ulm University]
{Institute of Electrochemistry, Ulm University, Albert-Einstein-Allee 47, 89081 Ulm, Germany}

\author{Timo Jacob}
\affiliation[Ulm University]
{Institute of Electrochemistry, Ulm University, Albert-Einstein-Allee 47, 89081 Ulm, Germany}
\alsoaffiliation[Helmholtz-Institute-Ulm (HIU)]
{Helmholtz-Institute-Ulm, Helmholtzstr. 11, 89081, Ulm, Germany}
\alsoaffiliation[Karlsruhe Institute of Technology (KIT)]
{Karlsruhe Institute of Technology (KIT), P.O. Box 3640, 76021, Karlsruhe, Germany}

\email{timo.jacob@uni-ulm.de}

\title[An \textsf{achemso} demo]
  {Unraveling the Optical Signatures of Polymeric Carbon Nitrides: Insights into Stacking-Induced Excitonic Transitions}

\begin{document}



\begin{abstract}
Two-dimensional (2D) materials have attracted considerable attention due to their unique physicochemical properties and significant potential in energy-related applications. 
Polymeric carbon nitrides (PCNs) with 2D stacked architecture show promise as photocatalysts for solar-to-fuel conversion and as versatile 2D semiconductors. However, the lack of a clear definition of the exact structural model of these materials limits our fundamental understanding of their unique properties.
Here, we investigate the structure-induced optical properties of PCNs through \textit{ab initio} calculations.
Our study on the electronic and optical properties of PCNs highlights the significant influence of structure on their behavior, especially near band edges. The analysis reveals that the degree of condensation and corrugation influences the electron/hole localization and the energy levels of $\pi$ electrons, which are crucial for the optical behavior. In addition, the microstructures of 2D configurations lead to divergent optical properties in 3D configurations, with characteristic peaks identified at 350 nm and interlayer interactions at 400--500 nm, depending on the specific microstructures. Through observations over 2D and 3D structures, we elucidate exciton photophysical processes in PCN materials.
This highlights the substantial differences in optical properties between actual 2D and 3D structures, while also demonstrating the potential for carrier and energy transport mechanisms to occur perpendicular to the plane. 
Finally, our results provide deep insights into the understanding of previously hidden microstructural, electronic, and optical properties of PCNs, paving the way for further performance and property enhancements in this class of materials.
\end{abstract}

\section{Introduction}
Quantum confinement effects of single-atom catalysts, 1D and 2D nanostructured materials have attracted considerable attention due to their unique physicochemical properties and significant potential in energy-related applications.\cite{novoselov2004electric,Wang2009,Mendelson2021,Wang2023}
Among them, intensive investigations have aimed to control the 2D layered structures and to harness the exceptional properties for practical applications.\cite{Wangn2009}
In particular, research on 2D semiconductor materials has received considerable interest in artificial photocatalysis due to the tunability of the electronic structure. \cite{duan2015two,an2022perspectives}

In particular, 2D backbone materials composed of heptazine and \textit{s}-triazine, collectively referred to as PCN, have gained significant attention due to their ease of synthesis and versatility in (photo-)catalytic performance and applications such as CO$_2$ conversion, oxygen evolution/reduction, hydrogen peroxide production, and hydrogen generation, all under UV and/or near-visible irradiation. \cite{Ma2020,Talapaneni2020,Krivtsov2020,Lin2020,Teng2021,Zhao2021,Xu2021,Ou2022,Wang2022,Zhang2022,Wu2022,Xie2023,Jia2023}

The detailed classification of PCN materials depends on their intricate structures and properties. Among these classifications, the conventional PCN structures such as melon and graphitic carbon nitride exhibit a suitable bandgap (2.7 eV), which can extend the light absorption to visible light\cite{inoki2020lowpressure, mitoraj2021study}.
Moreover, ionic carbon nitrides such as poly(heptazine imide) (PHI) have attracted attention owing to their unique properties, such as charge accumulation capabilities for the delayed catalysis\cite{Adler2021}, or very low photocurrent onset in photoelectrodes\cite{adler2021sol,pulignani2022rational}, showcasing the material's potential in numerous applications.

Despite the heightened interest in PCN materials, we still have a limited understanding of their photophysical and chemical properties and how these are connected to their structures. Therefore, it is difficult to explain their photochemical properties, which often depend heavily on synthesis conditions.\cite{Xie2023,Jia2023} Additionally, the relationship between charge carrier dynamics and photocatalytic efficiency is barely reported.\cite{Godin2017}

Computational simulations, particularly those utilizing \textit{ab initio} calculations, are a promising tool for understanding the atomic-level properties of materials, including their electronic structure. The calculations have been performed in order to identify the desirable model structure for PCN in order to reveal its unique properties. It is expected that PCN exhibits strong exciton binding, similar to other organic semiconductors, in a Frenkel-type excitation.\cite{Wei2013,Melissen2015,zhang2017optimizing}
However, due to the uncertainty of the structural models representing PCN and their calculated results differing from actual experiments, there is a need for better structural models that can accurately represent the properties of PCN.
In our previous studies, we calculated the thermochemical stability of reported PCN models and concluded that the actual PCN configuration is closer to a linear structure rather than a 2D structure, coexisting with melon and graphitic structures.
Furthermore, the energetic preference for stacking over size expansion allowed us to provide valuable insights into the shape of the real PCN structure.\cite{im2023structure}

In this study, we explore the intricate relationship between the structural properties of PCN and their impact on the optical behavior through \textit{ab initio} calculations.
Our research elucidates common characteristics inherent to both 2D and 3D PCN materials, notably the localization of electrons/holes and the significant role of $\pi$ electrons associated with carbon-nitrogen bonds. This study reveals that microstructural variations significantly impact the $\pi$ electron energy levels in 2D PCNs, thereby defining their optical signatures. Moreover, we have identified intricate interlayer interactions within these structures, which are integral to the unique configuration of each 2D layer.

\section{Results and discussion}

\subsection{Structural Models}

\begin{figure*}[htp]
\centering
\includegraphics[width=\textwidth]{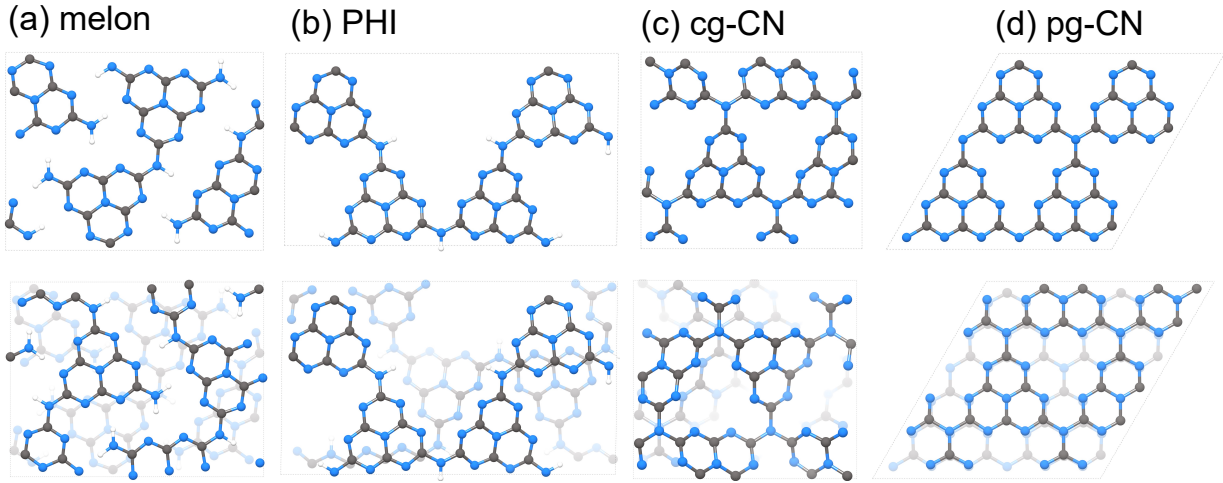}
    \caption{The illustration of the PCN models used in this study. The upper is the 2D monolayer and the lower is the stacked 3D structure.}
    \label{str}
\end{figure*}

The elucidation of PCN structures has been a recurring topic for a decade, beginning with the proposal of melon and graphitic carbon nitride structures.\cite{cao2015polymeric,zhao2015graphitic,liu2016graphitic,kessler2017functional}
The discovery of the PHI structure, also known as ionic carbon nitride, synthesized through the molten salt method, further contributed to the diversity of PCN architectures. \cite{Chen2017,Schlomberg2019}
However, the absence of the exact structure model for PCN still presents a challenge in establishing the structure--electronic (optical) relationship, impeding a thorough understanding.
Previous research has mainly focused on investigating the electronic and optical structures of 2D monolayers and their respective configurations.\cite{Wei2013, Steinmann2017, ReFiorentin2021}
In our previous research, we successfully established the thermodynamic conditions governing the PCN structural motifs under practical synthetic conditions. \cite{im2023structure}
This thermochemical stability suggests that the narrow and highly stacked heptazines locally possess melon (linear) and graphitic (fully condensed) regimes. 
The main objective of the present study is now to identify and characterize the optical-electronic properties depending on the structural features.

Figure \ref{str} illustrates the 2D monolayer and 3D stacked models of the heptazine-based carbon nitrides used in the present study, representing melon (linear structure), PHI, and both the corrugated (cg--CN) and planar graphitic carbon nitride (pg--CN) structures, respectively. 
While the graphitic structure (100\% degree of condensation) is regarded as "ideal," both theory and experiment suggest the partial presence of the graphitic domains within less condensed matrics.\cite{kessler2017functional,inoki2020lowpressure, SchulzeLammers2022,im2023structure,Xie112023}
This study excludes the effect of structural inhomogeneity, since the electronic contribution of the dangling triazine or heptazine moiety near the band edges is negligible.\cite{mitoraj2021study} 
Due to the robust exciton binding energy and a compact exciton radius of the Frenkel exciton, commonly assumed for heptazine-based carbon nitrides, the model structures employed in this study satisfactorily describe the localized electronic nature and sufficiently include the comprehensive properties of the general PCN structure. \cite{feierabend2019optical}

\begin{figure*}[htp]
\centering
\includegraphics[width=0.40\textwidth]{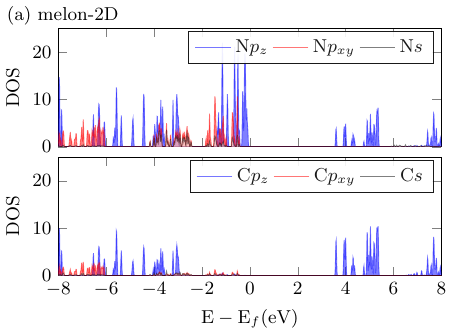}
\includegraphics[width=0.40\textwidth]{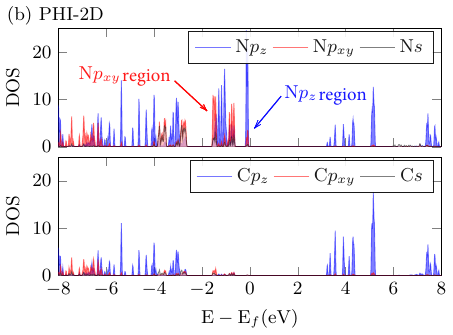}
\includegraphics[width=0.40\textwidth]{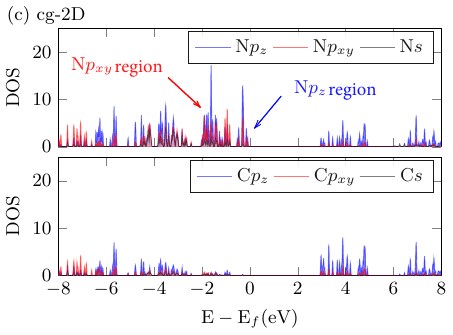}
\includegraphics[width=0.40\textwidth]{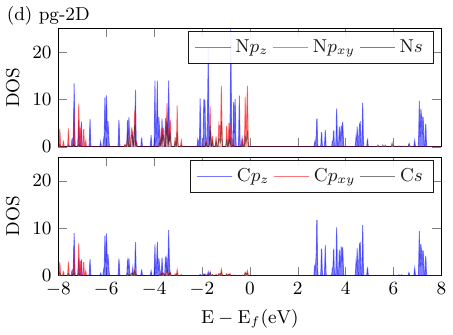}
\includegraphics[width=0.40\textwidth]{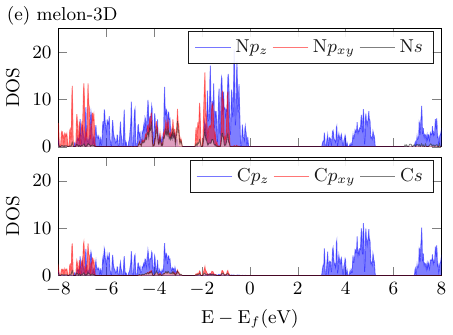}
\includegraphics[width=0.40\textwidth]{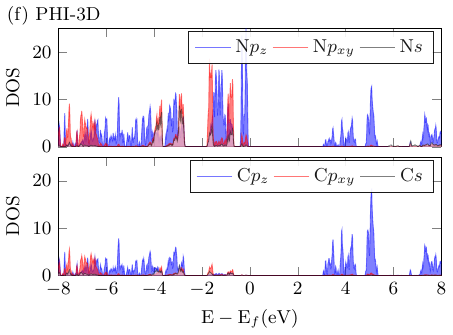}
\includegraphics[width=0.40\textwidth]{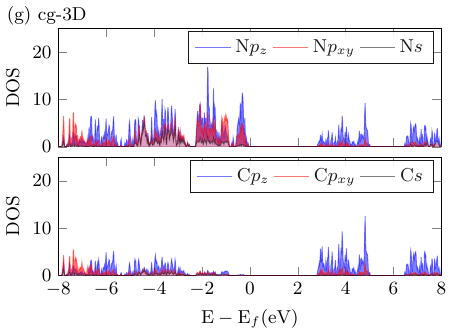}
\includegraphics[width=0.40\textwidth]{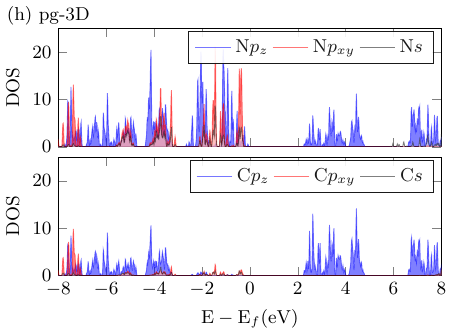}
    \caption{The projected density of states of (a, e) melon, (b, f) PHI, (c, g) pg--CN, and (d, h) cg--CN. The projection is performed on $s$, $p$ orbitals corresponding to carbon and nitrogen atoms. The trivial contribution of hydrogen is omitted. Fermi energy referenced to 0 eV.}
    \label{dos}
\end{figure*}

In Figure \ref{dos}, we analyzed the projected density of states (PDOS) of both 2D monolayer and 3D stacked structures using DFT with the HSE06 hybrid exchange-correlation functional. 
This analysis was conducted to characterize the orbital distribution of the elements, based on structural effects such as the degree of condensation, the degree of corrugation, and stacking.
It is commonly observed in both the 2D monolayer and 3D stacked models that nitrogen orbitals largely contribute to the occupied orbitals near the valence band maximum (VBM). 
The unoccupied orbitals near the conduction band minimum (CBM) are shared by both carbon and nitrogen atoms. 
This common configuration indicates that the typical electronic transition near the Fermi level in PCNs occurs between the occupied N-$p$ orbitals and the unoccupied C-$p$ and N-$p$ shared orbitals. 
This finding strongly supports the conclusion that $\pi$--$\pi^{*}$ transitions are the prevalent electronic transitions in PCN materials.\cite{An2021}
To provide more specific details, the PDOS analysis of both 2D and 3D structures reveals a significant alteration in the primary contribution of the VBM orbitals, shifting from N$p_{z}$ orbitals to N$p_{x+y}$ orbitals as the degree of condensation increases (melon $\rightarrow$ PHI $\rightarrow$ g-CN). 
Notably, the N$p_{x+y}$ orbital contribution becomes the most prominent in the graphitic and planar structures (pg--CN). 
In these structures, the projected electron density of N$p_{x+y}$ can represent both $\pi$ bonding and non-bonding electron pairs due to the distorted $p_{z}$ orientation caused by the buckled structure, which is not accounted for in simple spatial projections.
For clarity, we present the actual electron density corresponding to the respective energy level in Figure \ref{s1}. 
The electron density clearly illustrates that in flat structures (melon and pg--CN), the actual lone-pair electrons correspond to the N$p_{x+y}$ orbitals. 
In corrugated structures (PHI and cg--CN), the lone-pair electron density is limited to energy levels where the density of states of N$p_{x+y}$ is higher than that of $p_{z}$ (see Figure \ref{dos}, marked with arrows).
This lower energy profile of the lone-pair electrons in corrugated structures, approximately 0.5$-$1 eV lower compared to the $p_z$ orbitals, indicates that the utilization of lone-pair electrons is less favorable than that of $\pi$ electrons. 
The elevated energy levels of the lone-pair electrons in pg--CN can be attributed to the substantial confinement of electrons resulting from the highly symmetrical structure. \cite{Wei2013}
This symmetry enhances interactions among N$p_{x+y}$ orbitals, maintaining the locally stable structure. \cite{Gao2020}
We have also observed that the impact of stacking on the electronic structure differs between planar structures (melon and pg--CN) and buckled structures (PHI and cg--CN). 
In planar structures, the emergence of small N$p_{z}$ peaks at the VBM can be attributed to the overlapping $p_{z}$ orbitals between the layers (Figure \ref{dos}). 
These energy levels indicate the presence of interactions between the layers, occupying the band edge.
On the other hand, in buckled structures, the peaks in the stacked 3D structures exhibit almost the same pattern as the 2D monolayer, indicating minimized interaction between different layers. 
Importantly, both the overlapping $p_{z}$ orbitals are situated near the Fermi level.
This suggests that interlayer interactions in 3D structures have a significant influence on the bandgap.

\begin{figure*}[htp]
\centering
\includegraphics[width=0.23\textwidth]{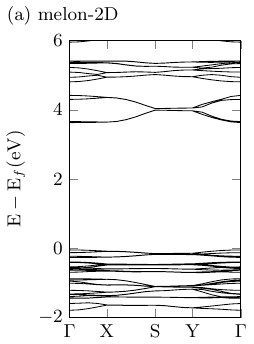}
\includegraphics[width=0.201\textwidth]{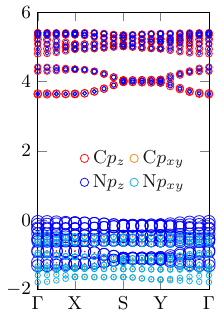}
\includegraphics[width=0.23\textwidth]{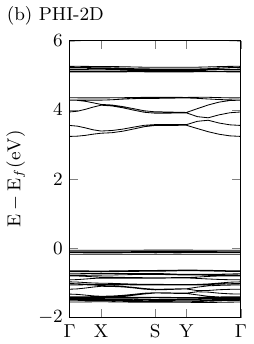}
\includegraphics[width=0.201\textwidth]{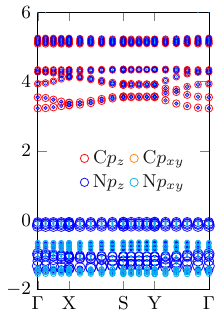}
\includegraphics[width=0.23\textwidth]{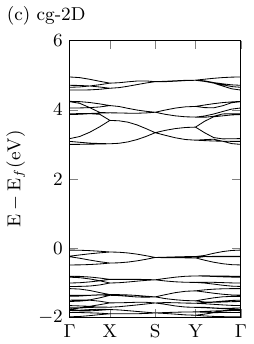}
\includegraphics[width=0.201\textwidth]{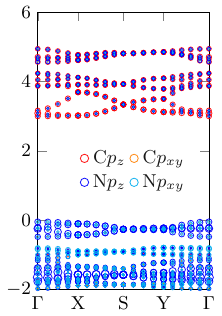}
\includegraphics[width=0.23\textwidth]{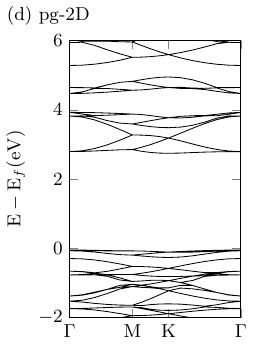}
\includegraphics[width=0.201\textwidth]{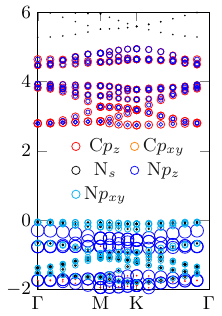}
    \caption{The band structure (left) and its orbital projection (right) of (a, e) melon-2D, (b, f) PHI-2D, (c, g) pg--CN-2D, and (d, h) cg--CN-2D. The projection is performed on $s$, $p$ orbitals corresponding to carbon and nitrogen atoms. The trivial contribution of hydrogen is omitted. Fermi energy is referenced to 0 eV.}
    \label{band2d}
\end{figure*}

\begin{figure*}[!h]
\centering
\includegraphics[width=0.23\textwidth]{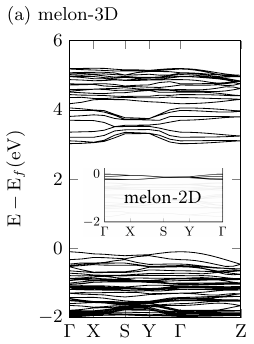}
\includegraphics[width=0.201\textwidth]{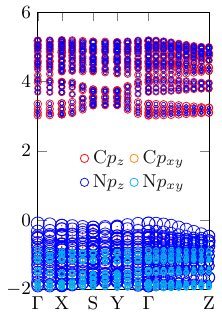}
\includegraphics[width=0.23\textwidth]{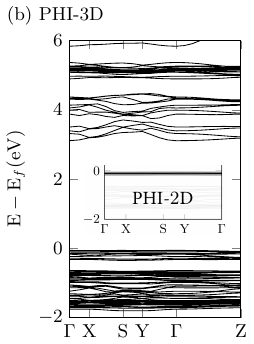}
\includegraphics[width=0.201\textwidth]{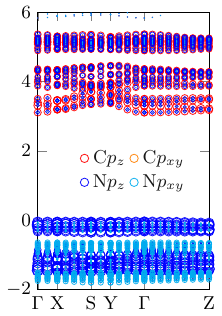}
\includegraphics[width=0.23\textwidth]{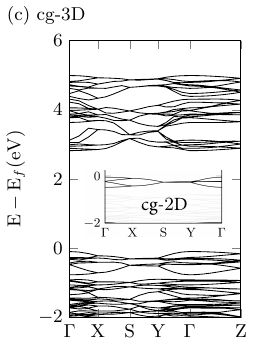}
\includegraphics[width=0.201\textwidth]{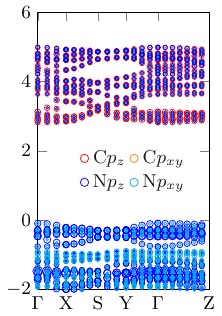}
\includegraphics[width=0.23\textwidth]{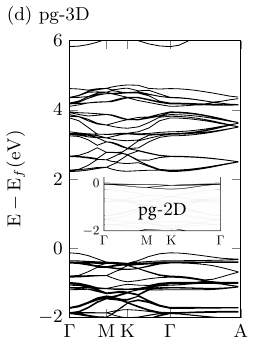}
\includegraphics[width=0.201\textwidth]{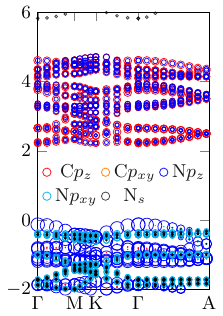}
    \caption{The band structure (left) and its orbital projection (right) of (a, e) melon-3D, (b, f) PHI-3D, (c, g) pg--CN-3D, and (d, h) cg--CN-3D. The projection is performed on $s$, $p$ orbitals corresponding to carbon and nitrogen atoms. The trivial contribution of hydrogen is omitted. Fermi energy is referenced to 0 eV. The inset provides a comparison of energy levels near the VBM in the corresponding 2D structure, illustrating different degrees of interlayer interactions.}
    \label{band3d}
\end{figure*}

In Figure \ref{band2d} and \ref{band3d}, we present the band structure and its elemental/orbital projection for both the monolayer and stacked structures, respectively. 
Firstly, a flat band structure is evident at the band edges and other energy levels, indicating localized electrons resulting from discontinuous potentials. \cite{Wei2013,ReFiorentin2021,Wei2018}
The pg--2D structure exclusively exhibits an indirect bandgap, whereas the other 2D and 3D structures have direct bandgaps. 
However, the disparity between the direct and indirect gap could be negligible not only in the pg--2D structure but also the others due to the observed flat band nature throughout the highly symmetric $k$-point.
These small deviations between the indirect and direct transitions may give rise to various shallow trapping levels due to the strongly localized electrons and holes.\cite{Jing2017,Ruan2020,Rieth2021,Pulignani2022}
The calculated bandgap energy of the 2D and 3D structures is provided in Table \ref{bandgap}. 

\begin{table}[!b]
\centering
\caption{The calculated bandgap (HSE06) of PCN structures.}
\begin{tabular}{@{}llll@{}}
\toprule
Structure  & bandgap (eV) & Structure  & bandgap (eV) \\ \midrule
melon-2D   & 3.68          & melon-3D   & 3.12          \\
PHI-2D     & 3.32          & PHI-3D     & 3.18          \\
cg-2D & 2.99          & cg-3D & 2.95          \\
pg-2D & 2.79          & pg-3D & 2.35          \\ \bottomrule
\end{tabular}
\label{bandgap}
\end{table}

The bandgap tends to decrease as the 2D monolayer evolves into the stacked 3D structure. 
This decrease can be attributed to the quantum size effect of confined electrons. \cite{balzani2014photochemistry}
Specifically, the bandgap reduction in PCNs is a consequence of the discrete arrangement of energy levels, resulting from the interplay of localized electrons within each layer. 
This interaction of localized electrons leads to a rearrangement of energy levels, the energy levels of occupied orbitals become relatively elevated while the unoccupied orbitals are lowered accordingly.
The bandgap reduction due to stacking is more pronounced in the planar structures (melon: 0.56 eV and pg--CN: 0.44 eV), whereas it is relatively suppressed in the corrugated structures (PHI: 0.14 eV and cg--CN: 0.04 eV). 
Hence, we can comprehend the role of corrugation in the electronic structure, which minimizes interlayer interaction by distorting the structure.

Particularly in the 3D stacked structures, we observe that as the degree of condensation increases, the structure becomes more corrugated, except for the pg--CN structure, which is a local minimum structure.
However, this pronounced corrugation does not lead to a substantial decrease in the bandgap. 
This suggests that for PCN structures where the degree of condensation\cite{im2023structure} exceeds that of PHI (75\%), the local flattening significantly influences the effective bandgap. 
In brief, our findings underscore the significance of interlayer interactions and depth-related $\pi$ electrons as the principal contributors to the bandgap of conventional PCN structures.
Furthermore, we emphasize that the calculated bandgap from the ground state should correspond to the positions of the absorption peaks in the absorption spectra, rather than the main peaks in the emission spectra at 2.7 eV or the bandgap obtained from the absorption coefficient using the Kubelka--Munk theory.\cite{kubelka1931article}
In calculating the bandgap of the periodic PCN models using \textit{ab initio} methods, it is crucial to strictly adhere to the Franck--Condon principle, due to the current absence of methods to obtain the excited geometry.
Therefore, determining the actual bandgap corresponding to the 2.7 eV emission necessitates further calculations that consider both the excited geometry and vibrational contributions.
This also indicates that the major peak position near the band edge from the experimentally observed absorption spectra (\textit{ca.} 400 nm, approximately 3.09 eV) should align with either the calculated bandgap or the position of the calculated absorption spectra.\cite{merschjann2013photophysics}
Our computed bandgaps for PCNs, which consistently group at approximately 3 eV for 3D structures, such as the linear melon and PHI configurations, align seamlessly with this interpretation (cf. Table.\ref{bandgap}). 
Further discussions on detailed optical--electronic characteristics depending on structures will be addressed in the following section.

\subsection{Dielectric Functions and Optical Properties}

\begin{figure}[htp]
\centering
\includegraphics[width=0.46\textwidth]{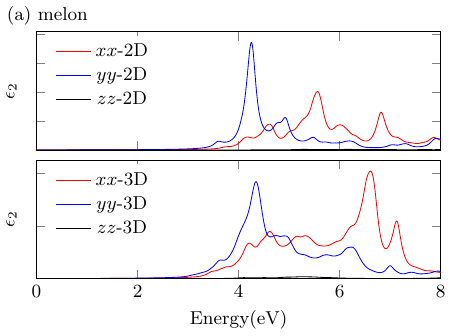}
\includegraphics[width=0.46\textwidth]{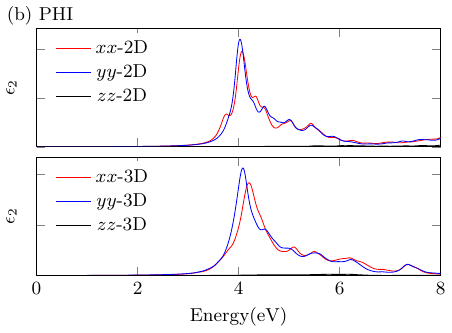}
\includegraphics[width=0.46\textwidth]{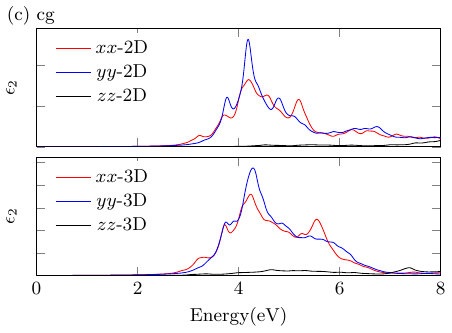}
\includegraphics[width=0.46\textwidth]{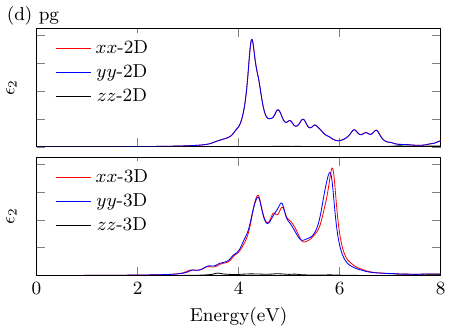}
    \caption{The imaginary part of the macroscopic dielectric functions on top of the HSE@Casida method. Imaginary parts corresponding to the respective cartesian direction ($xx, yy, zz$), obtained by Eq. (\ref{eq4}) are shown.}
    \label{dielec}
\end{figure}

In this section, we discuss the dielectric function and optical properties of both 2D and 3D PCN structures (see Figure \ref{dielec}).
Our primary emphasis is on the imaginary part of the dielectric function, which closely correlates with absorption spectra.
To establish the validity of our calculations, we compared the imaginary parts of the dielectric functions of pg--CN--2D between the TDHF@Casida method in our study and the GW@BSE method acquired from literature (Figure \ref{s2}). 
We observed systematic shifts between the TDHF@Casida and GW@BSE results, indicating the potential applicability of a scissor correction.
Thus, we have accordingly applied the scissor correction to other 2D structures to approximate the GW level of theory, which includes electron--hole interactions.
However, it is important to note that for 3D structures, the precision of the scissor correction is constrained by limitations in existing literature. \cite{ReFiorentin2021,Wei2013}
Nonetheless, the consistent and systematic shift in energy levels observed from 2D to 3D structures (both cg and pg) supports the continued validity of the scissor correction, even for the evaluated 3D structures.\cite{Guilhon2018,you2021nonadiabatic}

In the calculated imaginary parts, a prominent bright exciton emerges, consistently appearing near 4 eV in both 2D and 3D PCNs. 
Below 4 eV (310 nm), we observe multiple bright states with relatively weak intensities, which are related to structural variations within the PCNs.
Interestingly, when transitioning from 2D to 3D PCN structures, we discern an overall similarity of the imaginary parts, especially near the band edges.
While interlayer interactions induce distinct changes in the band structure, the changes in optical properties remain minor due to their relatively weaker intensities compared to the absorption peaks near 4 eV.
This observation illustrates that, despite the pronounced structural differences in PCN, the absorption behavior near the band edge can yield only minor variations.
Consequently, this explains the challenges of characterizing PCN microstructures from absorption spectra alone.
Figure \ref{dielec}(a) shows the imaginary part of dielectric functions for 2D and 3D melon structures, revealing distinct optical anisotropy. \cite{li2021ultrafast,giusto2022optical}
This distinctive feature reflects the linear structure characteristic of melon, raising considerations regarding its directional polarizability and potential limitations in optical efficiency.

\begin{figure}[htp]
\centering
\includegraphics[width=0.46\textwidth]{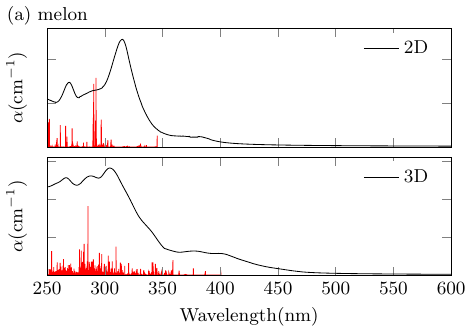}
\includegraphics[width=0.46\textwidth]{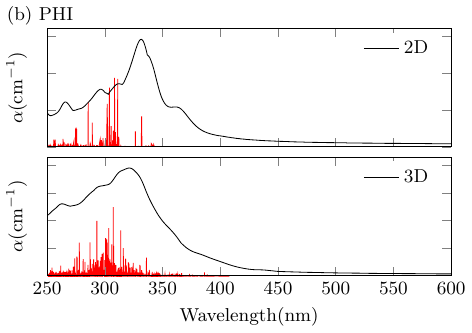}
\includegraphics[width=0.46\textwidth]{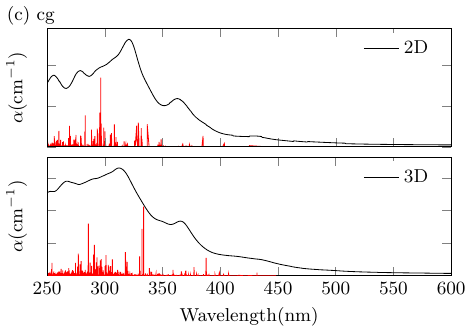}
\includegraphics[width=0.46\textwidth]{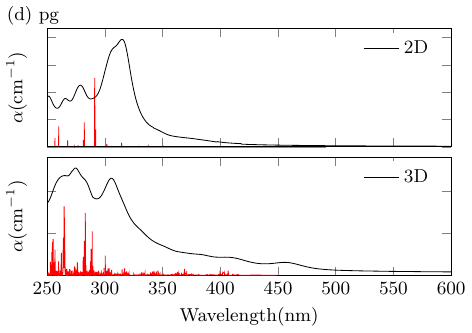}
\includegraphics[width=0.46\textwidth]{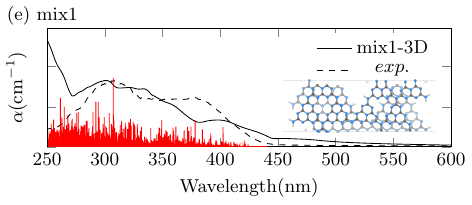}
\includegraphics[width=0.46\textwidth]{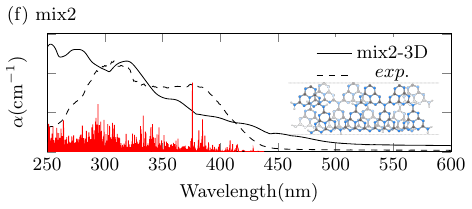}
    \caption{The absorption spectra and oscillator strength were calculated for (a) melon, (b) PHI, (c) cg, and (d) pg in both 2D and 3D structures. The absorption spectra and oscillator strength of (e) mix1-3D and (f) mix2-3D, which depict the amorphous structure, are presented alongside the experimental values which is a dashed line.}
    \label{abs}
\end{figure}

In Figure \ref{abs}, we present average absorption spectra polarized in the respective spatial direction for 2D and 3D PCNs.
We also provide the oscillator strength corresponding to the optical transition probability plotted with regard to wavelength.
Notably, a discrepancy between absorption spectra and oscillator strength indicates that calculated absorption encompasses transitions from different energy levels.\cite{ruger2015efficient}

This misalignment most likely contributes to the major disparity between the calculated absorption spectra of PCNs and experimental observations.
In line with the imaginary part of dielectric functions in Figure \ref{dielec}, strong oscillator strength consistently manifests around 300 nm (approximately 4.1 eV) in models of 2D and 3D structures. 
We found that this peak reflects the highest degenerate state of excited $\pi$ electrons in C--N bonds within PCN structures (cf. Figure \ref{s1}).
Additionally, the absorption peak and oscillator strength around 350 nm are linked to structural features.
The intensity of this peak is pronounced in the buckled structures (PHI, cg) and shifts across structures, correlating with the degree of condensation within the PCN.
Lastly, the absorption peak appearing around 400 nm (3.09 eV) is a consequence of interlayer interactions.
This wavelength for the interlayer interactions is in good agreement with the literature observing the interplanar electron hopping in melon from transient absorption spectroscopy and transient photoluminescence using the pump wavelength of 388 nm. \cite{merschjann2015}
The peak intensity and patterns are also in agreement with the aforementioned PDOS results regarding the variation in interlayer interactions due to corrugations.
Additionally, the absorption edge in each 2D structure undergoes a red shift in 3D structures.
This red shift caused by interlayer overlap results in the formation of new peaks in melon (400 nm), cg (400--450 nm), and pg (450--500 nm), but it is not observed in the PHI structure.\cite{Chen2021}
It is noteworthy that the subtle red shift observed in PHI can be attributed to the minimized interlayer interactions resulting from its structural flexibility. 
This structural adaptability allows the PHI structure to mitigate interactions, including interlayer interactions and the lone-pair repulsion of nitrogen atoms, by controlling interlayer spacing and local corrugation, respectively. 
Consequently, the PHI structure exhibits remarkable independence between layers and heptazine units, simultaneously.
The suggested independence between the different layers can be confirmed by the band structure in Figure \ref{band2d} and \ref{band3d}.
When the PHI structure becomes 3D (from 2D), the simple overlap of energy levels is exhibited near band edges, while in the corrugated structure, it can be seen that one energy level moves upwards, and the others shift downward.
It indicates that stronger interlayer interaction exists in the cg structure than in the case of the PHI structure. 
Additionally, the peaks ranging from 400--500 nm may provide the theoretical explanation that the forbidden transition, known as $n$--$\pi$ transition, can be attributed to the graphitic regions.
It is noted that these absorption peaks are specific to pure PCNs.
This study does not consider interactions with other chemical species, such as radicals, charged, or heteroatom species, which might alter the behavior.
The calculated absorption peaks for the PCN structural features are summarized in Table \ref{assignment}.

\begin{table}[]
\caption{The assignment of calculated absorption peaks for the PCN structural features.}
\begin{tabular}{@{}ccc@{}}
\toprule
Absorption feature & Attribution                 &         \\ \midrule
$\sim$300 nm       & degenerated $\pi$ electrons &         \\
around 350 nm      & structural feature          & \makecell{weak at melon and pg, \\ PHI (365 nm), cg (375 nm)} \\
around 400 nm      & interlayer interaction      & \makecell{\\ melon (400 nm),\\ cg (400-450 nm),\\ and pg (450-500 nm)}        \\ \bottomrule
\end{tabular}
\label{assignment}
\end{table}

In our prior research, we introduced two microstructure patterns for PCNs, representing more amorphous configurations.\cite{im2023structure}
Figure \ref{abs}(e) and (f) illustrate the amorphous structures (inset) and their absorption spectra and oscillator strength.
These two structures are virtual models that partially incorporate microstructures from melon (at the edge boundary), PHI, and graphitic structures. 
The mix1--3D structure contains minimal graphitic structures, positioned on both sides, while the mix2--3D structure features a larger graphitic region.
Although neither structure fully replicates the experimental absorption spectra, our analysis suggests that the proportion of graphitic regions in the mixed model structures is significantly lower than that in the actually measured structure.
However, the absorption edge extending to 450 or 500 nm can be easily found in the literature, due to the variety of synthetic conditions .\cite{Peng2018, Zhang112017}
Notably, by combining the oscillator strength from both structures, we can closely reproduce the absorption spectra observed in the experiment. 
This finding suggests the presence of stacking patterns within these two mixed structures, revealing the stacking patterns of the mixed microstructures.
These patterns are distinct from the typical AB stacking arrangements found in crystallites of the respective PCN microstructures.

\subsection{Interlayer Interactions: Influence on Electronic and Optical Properties}
\begin{figure*}[htp]
\centering
\includegraphics[width=\textwidth]{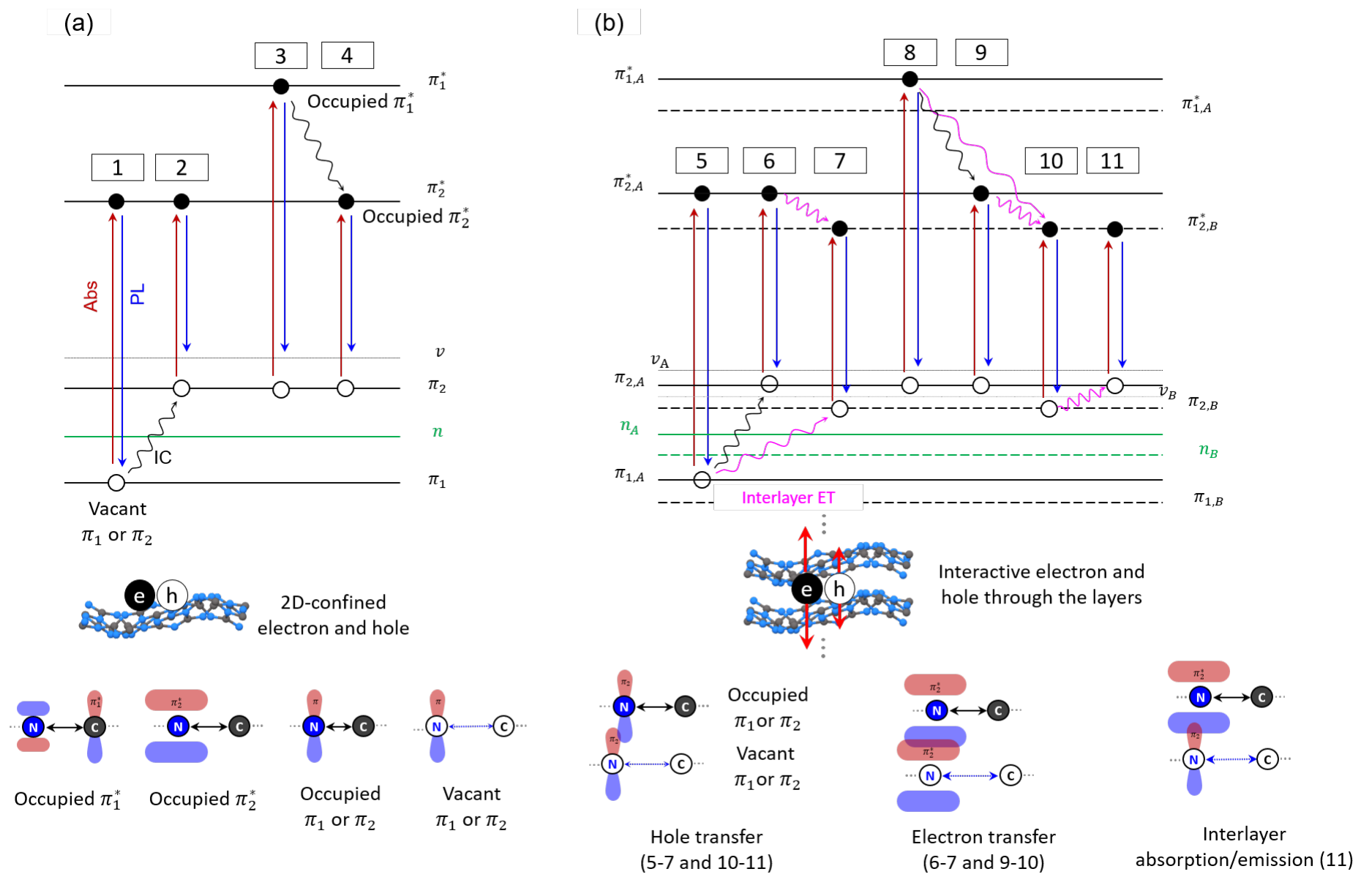 }
    \caption{The proposed photophysical process of the general PCN materials in (a) 2D monolayer and (b) 3D stacked structures. The electron density at the respective energy level is illustrated below. The red corresponds to the absorption process, the blue is the photoluminescence process, the black is the non-radiative internal conversion, and the magenta is the interlayer energy/electron transfer, respectively.}
    \label{mec}
\end{figure*}

Our results have consistently revealed the significant impact of interlayer interactions within PCN structures on their electronic structure and optical behavior near the band edge.
Additionally, we have established that fundamental features, common to both 2D and 3D PCN structures, including the degenerate $\pi$--$\pi^{*}$ transition peak associated with C--N bonds, the structural characteristic peaks dependent on the degree of condensation, and interlayer overlap peaks, are consistently present in typical PCN materials.
These observations indicate potential photophysical processes occurring in 2D and 3D PCNs, which are depicted in Figure \ref{mec}(a) and (b), respectively.
In 2D PCNs, the electronic structure of a monolayer near the band edges can be categorized into two $\pi$ states: $\pi_1$ (localized $\pi$ electron) and $\pi_2$ (delocalized $\pi$ electron), with a non-bonding (lone-pair) state positioned between them.
The bright excitation near 300 nm can be explained by $\pi$--$\pi^*$ transitions governed by the Franck--Condon principle for vertical transitions. 
Corresponding energy transitions can exist in two forms, Case 1 and Case 3 in Figure \ref{mec}(a), encompassing a wider energy range than the bandgap. 
We claim that excitons formed in this range represent hot carriers, supported by the fact that PCN absorption prominently occurs around 300 nm and 400 nm, while the primary emission converges at 450 nm.
However, generic 2D PCN structures have limitations in explaining the discrepancy between the absorption and emission profiles of PCNs.
This limitation occurs due to the absence of factors facilitating the relaxation of hot carriers. 
Instead, this energy range actually corresponds to bound exciton states, which enhance hot emission.
This indicates that the 2D PCN structure can show not only the direct hot emissions (Case 1 and Case 3) but also the emission after the cooling of hot excitons (Case 2 and Case 4) simultaneously.\cite{balzani2014photochemistry,Wang2017}
This discrepancy implies that a single layer of PCN can exhibit substantially different exciton behavior when compared to multilayered PCNs. \cite{qiu2021signatures}

Consequently, the reported absorption patterns of PCNs so far suggest the presence of PCNs in a multilayered form. 
This is consistent with our previous study, which suggested that stacking processes are thermodynamically favored over graphitic processes during the condensation reaction of PCN precursors.\cite{im2023structure}
In the stacked 3D PCN structure, a more discrete energy structure can emerge, owing to the interplay between the 2D layers and the alignment of energy levels (Figure \ref{mec}(b)). 
This characteristic indicates that the interlayer interactions within 3D structures can give rise to more complex excitonic processes. 
The proposed absorption profile exhibits a similar pattern to the 2D materials, with minor energy shifts attributed to the independence of each layer.
Notably, in 3D PCN structures, the cooling processes of hot carriers can gain prominence, derived by the overlaid $\pi$ orbitals from the stacking.
The orbital shapes corresponding to the respective energy levels are represented in Figure \ref{mec}(a).
In the absorption spectra, the electrons and holes generated around 300 nm can undergo internal conversion more readily in the stacked 3D structure compared to the 2D structure.
This phenomenon is attributed to the stabilization of the generated holes through Coulombic interactions with electron orbitals in other layers.
Consequently, Auger relaxation (Case 6) and the relaxation of excited electrons due to screened holes (Case 9) can be boosted in 3D structures. \cite{balzani2014photochemistry}
These factors offer fundamental insights into why PCNs exhibit multiple absorption peaks while displaying a prominent emission peak. 
Interlayer interactions are expected to lead to lower exciton binding energy in 3D structures compared to 2D counterparts, although a precise determination of exciton binding energy in 3D PCN structures necessitates further taking exact electron--hole interactions into account \textit{e.g.}, GW@BSE.

Based on the results so far, we assert that interlayer interactions play a pivotal role in the band edge properties of real PCN materials. 
Our calculations, in particular, highlight that the stacking indeed enables excited electrons, holes, and energy transport perpendicular to the layers.
Thus, it is anticipated that tuning the interlayer interactions, such as modifying interlayer distances by introducing alkali ions, could indeed influence the exciton dynamics and also the related catalytic activity.
This is supported by the fact that the generated electron/hole has sufficient orbital overlap with adjacent layers, and the interlayer spacing for electron transfer falls within the range of 3.2--3.4 \AA, suggesting the potential for Förster or Dexter-type energy transfer, in accordance with Fermi's golden rule, as well as electron transfer. \cite{balzani2014photochemistry}

Furthermore, the 3D band structure in Figure \ref{band3d} reveals that the energy at $\Gamma$--$Z$ or $\Gamma$--$A$ directions (both are perpendicular to the plane) requires only 0.1--0.2 eV higher energy than the direct bandgap. 
These small energy barriers can be less meaningful if the cooling process of hot excitons is actually more pronounced in 3D structures.
Significantly, this overlap of vertical orbitals selectively occurs at lone-pair N atoms in PCNs, implying that interlayer transitions can occur more effectively in the structures showing highly ordered crystallinity, rather than in strongly amorphous PCNs.
This could result in a slight blue shift in absorption/emission peaks.
Additionally, the possibility of such interlayer interactions provides a basis for explaining cascade phenomena, electronic or ionic conductivity, recombination, and charge accumulation (particularly in the presence of cations) in PCNs. \cite{Corp2017,Kroger2022}
For a more meticulous understanding of the electronic structure and corresponding optical properties in 3D structures, further studies at a higher theoretical level are warranted, encompassing electron--hole interactions.
However, our present studies already underscore the importance of interlayer interactions in PCNs and opens up avenues for future investigations for this intriguing class of materials.

\section{Conclusion}
In our study, we conducted \textit{ab initio} DFT calculations to unravel the complex relationship between the microstructural attributes of PCN materials and their optical behavior. This methodological approach allowed us to uncover distinctive electronic properties, underscoring the influence of condensation and corrugation degrees within PCN microstructures on their electronic and optical characteristics. Our analysis distinctly highlighted the correlation between absorption spectrum peaks and microstructural nuances, alongside the critical role of interlayer interactions in shaping these properties.
Notably, our findings illuminate the stark differences in optical behavior between 2D and 3D PCN structures, pointing to the pivotal role of interlayer interactions in modulating properties near the band edges. This research underscores the presence of localized electrons and the predominance of $\pi$ electrons, associated with carbon-nitrogen bonds, as fundamental determinants of optical properties near the band edge in both 2D and 3D PCNs.
The results of this study underscore the significant influence of interlayer interactions on the near band edge properties of generic PCNs, which paves the way for manipulating the electronic and/or optical properties as well as charge dynamics in PCNs through rational microstructural engineering.

\section{Experimental and computational methods}
\subsection{Computational details}
Calculations were conducted using the Vienna \textit{Ab initio} Simulation Package (VASP) version 6.3.2, which utilizes the projector-augmented wave (PAW) method.\cite{kresse1993ab,kresse1994ab,kresse1996efficiency,kresse1996efficient,kresse1999ultrasoft}
A plane wave energy cut-off of 400 eV was employed, and the wavefunction was optimized to an accuracy of $10^{-6}$ eV. Atomic coordinates were relaxed until the forces reached below $5 \times 10^{-2}$ eV/\r{A}. 
Gaussian-type finite-temperature smearing with a width of 0.01 eV was applied.
DFT-D3 dispersion correction was utilized to account for long-range interactions.\cite{grimme2010consistent,grimme2011effect}
The atomic and cell coordinates were relaxed using the Perdew, Burke, and Ernzerhof (PBE) exchange--correlation functional within the generalized gradient approximation (GGA).\cite{perdew1996generalized}
Accurate final energy results were obtained by performing single-point calculations on the PBE-optimized structures using the Heyd--Scuseria--Ernzerhof (HSE06) hybrid functional, which includes 25\% exact exchange and a screening factor of 0.2 \r{A}$^{-1}$.\cite{krukau2006influence}
The HSE06 functional was found to provide a good description of the electronic properties of PCNs.
For the monolayer, a vacuum layer with a thickness of 20 \r{A} was added to avoid interactions between the periodic images, and dipole correction was applied. The Brillouin zone integration for periodic models was performed using Gamma-centered \textit{k} grids.
The excitonic effects were calculated including the frequency-dependent dielectric functions and the oscillator strength by the time-dependent Hartree--Fock (TDHF) implanted in VASP.\cite{albrecht1998ab,rohlfing1998electron} The details are described in the Supporting Information and all data used in this study is available via Zenodo.\cite{im_2024_10844461}

\subsection{Sample preparation and diffuse reflectance spectroscopy}
The conventional yellow coloured polymeric carbon nitride was prepared by thermal polycondensation of 30 g melamine at 530 $^{\circ}$C for 4 h in a lid-covered crucible. Diffuse reflectance UV--vis spectra of solid was recorded using a Shimadzu UV2600 UV--vis spectrophotometer.

\begin{acknowledgement}

This work was funded by the Deutsche Forschungsgemeinschaft (DFG — German Research Foundation) through TRR 234 CataLight (project no. 364549901) as well as JA 1072/27-1 and BE 5102/5-1 (project no. 428764269). R.B. acknowledges funding from the European Union's Horizon Europe programme for research and innovation under grant agreement No. 101122061 (SUNGATE). The authors acknowledge support by the state of Baden-Württemberg through bwHPC and the German Research Foundation (DFG) through grant no INST 40/575-1 FUGG (JUSTUS 2 cluster).
 C.I. acknowledges the German Academic Exchange Service (DAAD, Ref. No. 91676720). 

\end{acknowledgement}

\begin{suppinfo}
\subsection{Supplementary methods}
The optical excitation energy is obtained from the transition matrix within the adiabatic linear response theory:\cite{furche2002adiabatic}
\begin{equation} \label{eq1}
\begin{bmatrix}
A & B \\
-B^* & -A^* \\
\end{bmatrix}
\begin{bmatrix}
\textbf{X}_s  \\
\textbf{Y}_s  \\
\end{bmatrix}
= E_s
\begin{bmatrix}
\textbf{X}_s  \\
\textbf{Y}_s  \\
\end{bmatrix}
\end{equation}
where $E_s$ is the excitation eigenvalues, $X_s$ and $Y_s$ are corresponding wavefunctions, $A$ and $-A^*$ are the resonant transition and antiresonant transitions from occupied orbitals to unoccupied orbitals, and $B$ and $-B^*$ are the coupling between the excitations and de-excitations, respectively.
In the HSE@Casida formulation, the matrices A:
$
A_{vc}^{v'c'} = (\varepsilon_v^{\rm HSE}-\varepsilon_c^{\rm HSE}v)\delta_{vv'}\delta_{cc'} + \langle cv'|V|vc'\rangle - \langle cv'|f_{\rm xc}|c'v\rangle~
$
, and the matrices B:$
B_{vc}^{v'c'} = \langle vv'|V|cc'\rangle - \langle vv'|f_{\rm xc}|c'c\rangle
$
, where the occupied $v,v'$ and unoccupied $c,c'$ states, include that interactions between electrons and holes are described by an effective nonlocal frequency-dependent kernel $f_{\rm xc}$. \cite{paier2008dielectric}
The Tamm-Dancoff approximation (TDA) neglects the off-diagonal coupling elements. Hence, the Eq. (\ref{eq1}) reduces to $AX_s=E_s X_s$.\cite{sander2015beyond}
The macroscopic dielectric matrix $\epsilon_M(\mathbf{q},\omega)$ is obtained by an inversion of the full microscopic dielectric matrix in giving long--wavelength limit $\textbf{q}\rightarrow 0$. It gives $\epsilon_M(\mathbf{q},\omega)=(\epsilon_{0,0}^{-1}(\mathbf{q},\omega))^{-1}$. Exploiting the TDA approximation, the solution of HSE@Casida ($E_s$ and $X_s$) can be used to obtain the macroscopic dielectric function:
\begin{equation} \label{eq2}
\epsilon_M^{\alpha\alpha}(\mathbf{q},\omega)=1+\frac{2}{\Omega}\frac{e^2\hbar^2}{\epsilon_0 m^2_0}\sum_{s} \left(
\left|\sum\limits_{c,v,\textbf{k},m}\frac{\langle c\mathbf{k}m|p_{\alpha}|v\mathbf{k}m\rangle}{\epsilon^m_{c\textbf{k}}-\epsilon^m_{v\textbf{k}}}^* X_s^{c,v,\textbf{k}}\right|^2
\times
\sum_{\beta=\pm1}\frac{1}{E_s - \beta\hbar(\omega+i\eta)} \right),    
\end{equation} 
where $p$ is the momentum operator with Cartesian coordinates $\alpha$, $m$ is z-component quantum number, $\Omega$ is the lattice volume, and $\eta$ is an infinitesimal number related to the exciton lifetime.\cite{gajdovs2006linear,rodl2008ab}
The oscillator strength $f^s_{\alpha}$ of state $s$ is given by
\begin{equation} \label{eq3}
f^s_{\alpha}\propto E_s\left|\sum\limits_{c,v,\textbf{k},m}\frac{\langle c\mathbf{k}m|p_{\alpha}|v\mathbf{k}m\rangle}{\epsilon^m_{c\textbf{k}}-\epsilon^m_{v\textbf{k}}}^* X_s^{c,v,\textbf{k}}\right|^2.    
\end{equation}
The imaginary part of the dielectric function for 3$\times$3 Cartesian tensor is given by 
\begin{equation} \label{eq4}
\epsilon_M^{2,\alpha\alpha}(\omega)=\frac{4\pi^2e^2}{\Omega}\lim_{q\rightarrow0}\frac{1}{q^2}\sum\limits_{c,v,\textbf{k},m} 2w_{\textbf{k}}\delta(\varepsilon_{c\textbf{k}+\textbf{q}}-\varepsilon_{v\textbf{k}}-\omega)
\times
\langle u_{c\textbf{k}+q\textbf{e}_{\alpha}}|u_{v\textbf{k}}\rangle
\langle u_{v\textbf{k}}| u_{c\textbf{k}+q\textbf{e}_{\alpha}}\rangle.
\end{equation}
, where the $\textbf{e}_\alpha$ is unit vectors for cartesian coordinates, $w_{\textbf{k}}$ is \textit{k}-point weights, and \textit{u} is the Bloch vector, respectively.
The real part of the dielectric tensor $\epsilon_M^{\text{real},\alpha\alpha}(\omega)$ is obtained by Kramers--Kronig relations
\begin{equation} \label{eq5}
\epsilon_M^{\text{real},\alpha\alpha}(\omega)=1+\frac{2}{\pi}P \int_{0}^{\infty} \frac{\epsilon_M^{\text{img},\alpha\alpha}(\omega')\omega'}{\omega'^2-\omega^2} \,d\omega ~.
\end{equation}
From the real and the imaginary parts of dielectric functions, the absorption coefficient is calculated.\cite{papadopoulos1991optical}

\newpage
\subsection{Supplementary data}
\setcounter{figure}{0}
\setcounter{table}{0}
\renewcommand{\thefigure}{S\arabic{figure}}
\renewcommand{\thetable}{S\arabic{table}}

\begin{figure*}[htp]
\centering
\includegraphics[width=\textwidth]{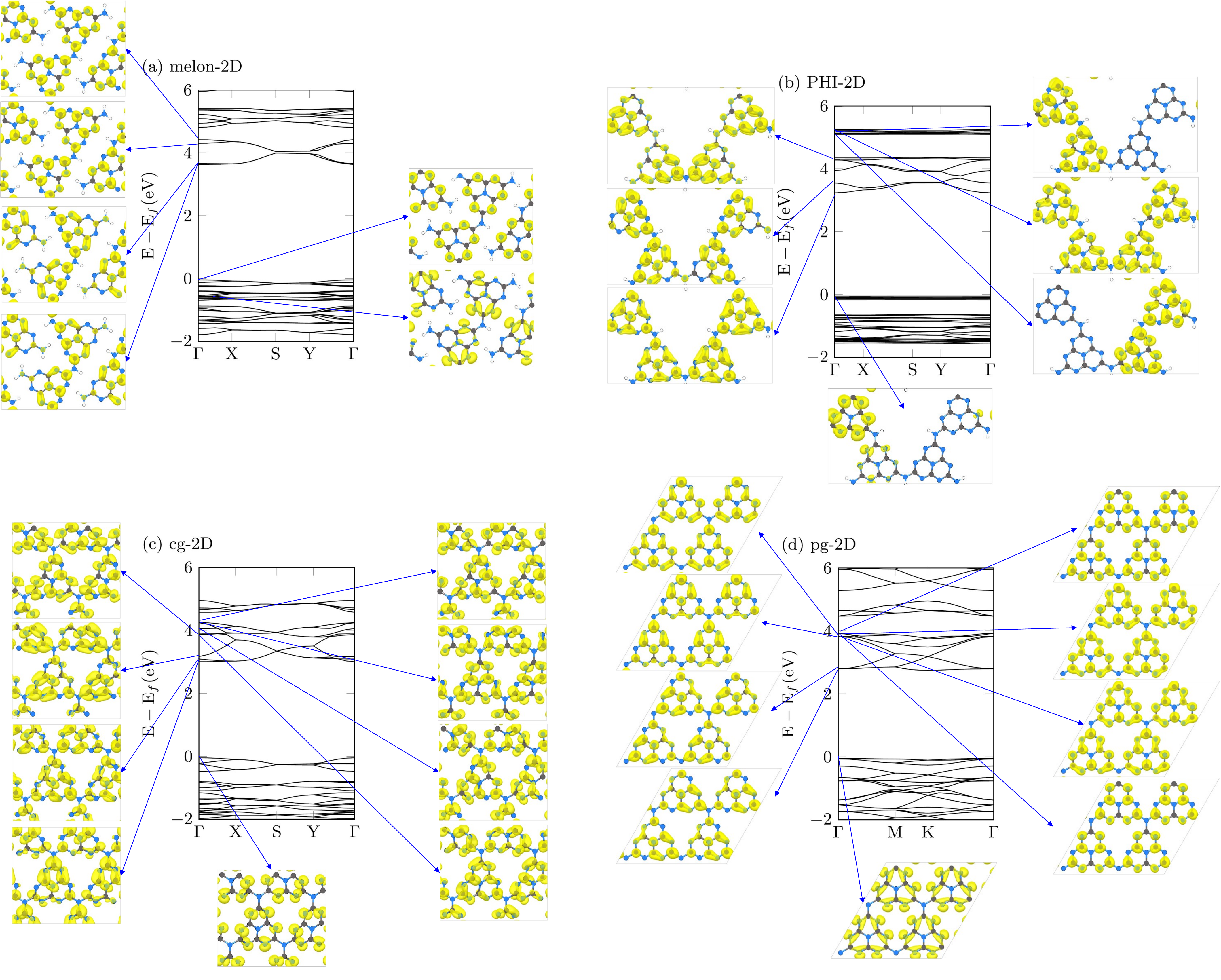}
    \caption{The depiction of electron density in (a) melon, (b) PHI, (c) cg, and (d) pg 2D structures corresponding to the respective energy level at $\Gamma$.}
    \label{s1}
\end{figure*}

\begin{figure*}[htp]
\centering
\includegraphics[width=0.7\textwidth]{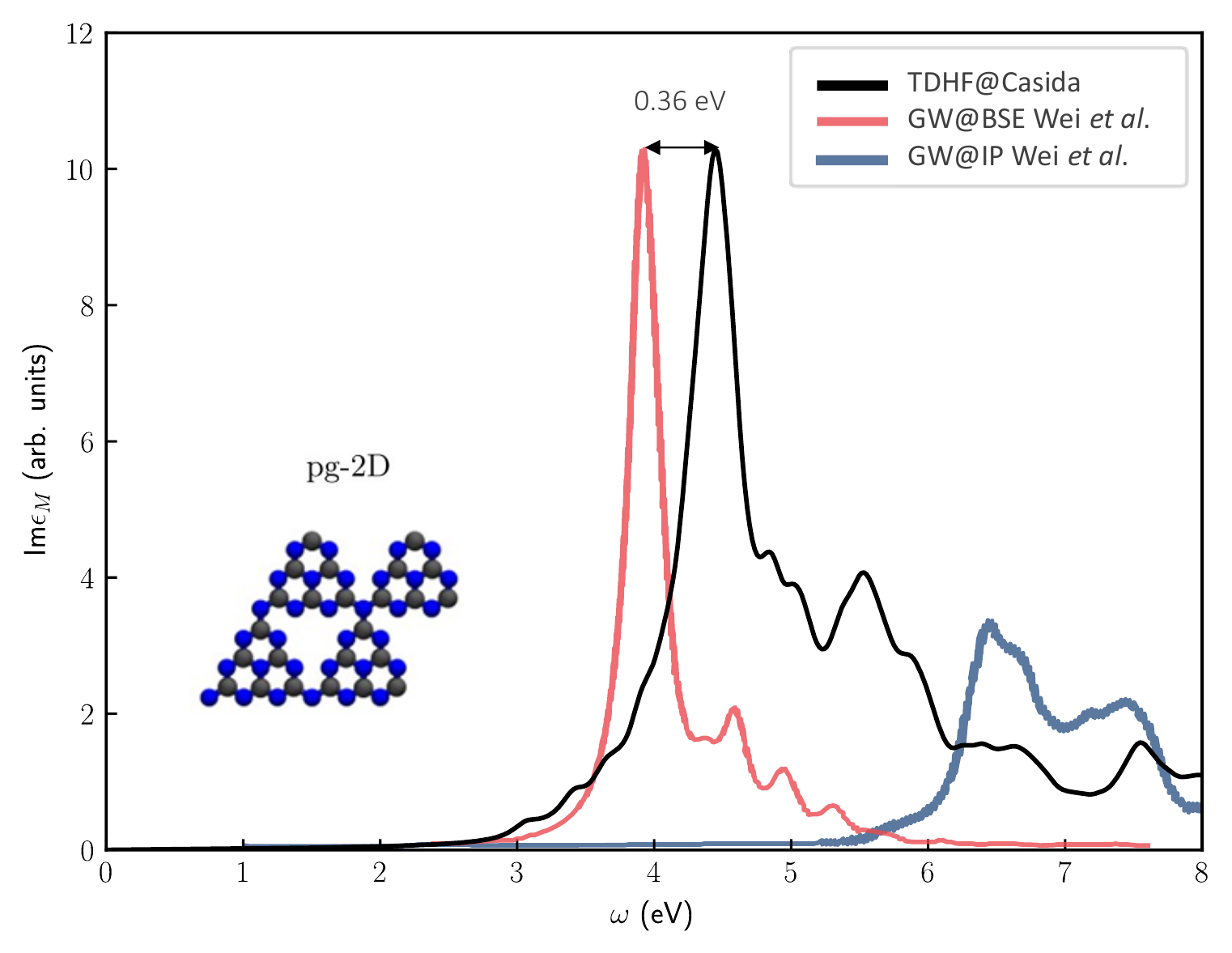}
    \caption{The calculated imaginary part of the dielectric function for pg-2D structure using TD@Casida (black), GW@BSE (red)\cite{Wei2013}, and GW@IP\cite{Wei2013} (blue) calculations. The scissor correction is applied based on the brightest state, which is commonly located around 4 eV.}
    \label{s2}
\end{figure*}

\newpage
\end{suppinfo}

\bibliography{achemso}

\end{document}